\documentclass[preprint]{aastex}       

\usepackage{psfig}

\newcommand{\APO}       {Apache Point Observatory}
\newcommand{\UCLan}     {University of Central Lancashire}

\newcommand{\LCO}       {Las Campanas Observatory}
\newcommand{\eg}        {e.g.}
\newcommand{\eqn}       {Eqn.}
\newcommand{\etal}      {et al.}
\newcommand{\ie}        {i.e.}
\newcommand{\fig}       {Fig.}
\newcommand{\figs}      {Figs.}
\newcommand{\kmsmpc}    {\mbox{~km~s$^{-1}$~Mpc$^{-1}$}}
\newcommand{\OmegaA}    {\mbox{$\Omega_{\rm A}$}}
\newcommand{\OmegaF}    {\mbox{$\Omega_{\rm F}$}}
\newcommand{\rah}       {\mbox{$^{\rm h}$}}
\newcommand{\rn}        {\mbox{$r_{\rm n}$}}
\newcommand{\rp}        {\mbox{$r_{\rm p}$}}
\newcommand{\Rlz}       {\mbox{$R(\lambda, z)$}}        
\newcommand{\teleang}   {\mbox{$\theta_{\rm T}$}}       
\newcommand{\rms}       {r.m.s.}
\newcommand{\WA}        {\mbox{$W_{\rm A}$}}
\newcommand{\WF}        {\mbox{$W_{\rm F}$}}

\begin{document}

\shortauthors{Newman} 
\shorttitle{Positioning errors in fiber spectrographs}

\title{Positioning errors and efficiency in fiber spectrographs}

\author{Peter R. Newman}

\affil{\APO, P.O.\ Box 59, Sunspot, NM 88349-0059; and Department 
of Astronomy, New Mexico State University, P.O.\ Box 30001, Las 
Cruces, NM 88003-8001} 

\email{prn@apo.nmsu.edu}

\begin{abstract}
The use of wide-field multi-object fiber-input spectrographs for 
large redshift surveys introduces the possibility of variations 
in the observed signal-to-noise ratio across the survey area due 
to errors in positioning the fibers with respect to the target 
image positions, leading to position-dependent errors in the 
survey catalog.  This paper brings together a comprehensive 
description of the sources of fiber-to-image position errors in 
different instrument designs, and quantifies their effects on the 
efficiency with which signal is recorded.  For point sources, a 
function relating a fractional efficiency and an equivalent 
aperture correction to the fiber-to-image position error, the 
fiber diameter and the image size is plotted for typical values 
of fiber and image sizes found in current instruments.  The tools 
required by observers to maximize the efficiency of 
fiber-spectrographic surveys are discussed.  
\end{abstract}


\section{Introduction}

The SDSS \citep{yorketal2000} and 2dF surveys 
\citep{boyleetal2001} exemplify the deep, wide-field redshift 
surveys now being conducted.  These surveys aim to produce 
homogeneous three-dimensional maps of $\sim3\times10^5$ to $10^6$ 
extragalactic objects to $B\simeq21$, about 10~per cent of which 
will be quasars.  Such surveys rely on multiplexed instruments to 
record spectra at rates of thousands of candidates per night.  
The multi-object optical fiber-input spectrographs being used for 
the 2dF and SDSS surveys represent a culmination of more than two 
decades of instrument development, during which time more than 35 
fiber spectrographs have been built or proposed, and more of 
which can be expected in the future.  

Constructing catalogs from such large observational data sets 
requires automated analysis if reliable and consistent spectral 
classifications and redshift measurements are to be achieved.  
The probability of correct results from automated analysis 
depends primarily on the signal-to-noise ratio (SNR) of the 
reduced spectra, and the SNR is strongly affected by how much of 
each target image actually falls on a spectrograph aperture.  
With single-object spectrographs, it is common practice to adjust 
telescope pointing during each exposure to maximize the signal 
recorded.  While difficult to achieve to any degree with 
multi-object instruments, such adjustments, at best, result in an 
optimal {\em mean\/} signal, averaged over all the apertures.  
Offsets between the fiber and image positions during exposure 
will clearly reduce the recorded signal for all types of observed 
object.  However, while typical galaxy images are $\ga 3$~arcsec 
across at all redshifts (for a Hubble constant of $75$\kmsmpc), 
giving some natural tolerance to fiber positioning errors, quasar 
images are usually seeing-limited at $\sim0.5$--$2.0$~arcsec 
diameter and thus demand greater precision in positioning each 
fiber.  

Where a survey is constructed from the consolidation of tiled 
multi-object spectrograph observations, each exposure will be 
made under different conditions, leading to variations in the 
rate at which signal is recorded.  In addition to random 
variations in fiber position errors, the errors may vary 
systematically across each tile leading to position-dependent 
selection effects that repeat from field to field.  Survey 
catalogs may thus suffer from very complicated position-dependent 
measurement errors if the variations in SNR are left uncorrected, 
creating problems for the analysis of large-scale structure.  

With perfect fiber positioning, telescope guidance, seeing 
conditions, and sufficiently wide fibers, essentially all of the 
incident image of an observational candidate will fall into the 
spectrograph input aperture.  In practice, spectrograph designs 
are such that the optimum fiber diameter is a compromise 
influenced by factors that include the telescope focal ratio, the 
fiber acceptance angle, the expected seeing disk size and the 
desire to minimize contamination of the object spectra with sky 
light \citep{brodielb88}, leaving many possible sources of loss 
of efficiency.  Mismatches between the telescope beam, fiber 
aperture and the central obstruction in reflecting telescopes may 
lead to systematic losses \citep{wynne92}.  Fibers are mounted on 
surfaces that only approximate the true telescope focal surface 
and telecentric alignment.  Atmospheric dispersion and 
differential refraction place strong constraints on fiber 
positioning \citep{donnellyetal89} and the timing of 
observations.  The accuracy of pointing and focus in modern 
telescopes is good, but still not perfect.  All of these factors 
lead to the conclusion that, in practice, some offset of the each 
fiber from its target image is unavoidable.  

How do these fiber positioning inaccuracies and other 
observational parameters affect the level of signal recorded?  
\citet{hill93} gave a qualitative introduction to most of the 
issues involved, and \citet{cuby94}\ addressed some of the issues 
quantitatively for the specific example of the VLT, but there has 
been little consolidation of the general problem in the 
literature.  The aim of this paper, then, is to provide a 
comprehensive list of the sources of fiber position errors and to 
quantify their effects on efficiency when fiber spectrographs are 
used to observe point sources such as quasars.  The rest of the 
paper is organized as follows.  Section 2 defines the functions 
that measure the efficiency of spectrographic observations in 
terms of fiber position error.  Section 3 describes the sources 
of position errors.  Section 4 quantifies the efficiencies 
achievable for different fiber sizes, seeing conditions and fiber 
position errors.  Finally, section 5 discusses the effects of the 
achieved efficiencies on wide-field surveys and the requirements 
for software tools to aid in their efficient execution.

\section{Efficiency of spectrographic observations}

The efficiency of observation, $\eta$, for an individual object 
targeted by a spectrograph may be defined as the fraction of the 
total light in the object's image that actually enters the 
spectrograph input aperture during an exposure.  For a planar 
aperture, this is given in general by
\begin{equation}\label{eq:eta_point_source}
 \eta = \frac{\int_{t_e}\int_{A(\mathbf{r})}\OmegaA\,dA\,dt}
             {\int_{t_e}\OmegaF\,dt}
\end{equation}
where \OmegaF\ is the intensity distribution of the image in the 
focal surface of the telescope, \OmegaA\ is the intensity 
distribution of the image in the plane of the aperture, $t_e$ is 
the exposure time, $A$ is the aperture and $\mathbf{r}$ is the 
separation vector from the center of the aperture to the center 
of the image in the plane of the aperture.  $\mathbf{r}$, 
\OmegaA\ and \OmegaF\ will all vary with time.  An aperture 
correction or equivalent loss in magnitudes may also be defined 
as $m_a = -2.5\log(\eta)$.  

The form of both \OmegaF\ and \OmegaA\ clearly depends on the 
nature of the objects being observed, the atmospheric seeing 
conditions and the telescope optical design.  This paper 
considers point sources as a worst case because the effect on 
$\eta$ of fiber position errors orthogonal to the optical axis 
will clearly be larger for point sources than for extended 
sources.  The ideal point-spread function is the so-called Airy 
pattern \citep[see for example][p.\ 180]{schroeder87}, but the 
detailed profile of the image of a point source after it has been 
convolved with telescope distortions, random motions from guiding 
errors and the effects of the atmosphere and then integrated over 
exposure times long enough to blur individual motions is, in 
practice, quite complex \citep{king71,woolf82}.  However, for the 
purposes of this paper both \OmegaF\ and \OmegaA\ for a point 
source image are adequately modelled by a 2-dimensional Gaussian 
with circular cross-section and known full width at half maximum 
(FWHM), $W$, given by
\begin{equation}\label{eq:mu_point_source}
  \Omega(\rho,\sigma) = \frac{1}{2\pi\sigma^2}
   \exp\left[\frac{-\rho^2}{2\sigma^2}\right]
\end{equation}
where $\rho$ is the radial position with respect to the center of 
the image, $\sigma = W/2\sqrt{\ln4}$, which may be expected to 
vary with time, and the constant factor normalizes 
$\int_{0}^{\infty}\Omega\,d\rho = 1$.  

It is well known that the atmosphere disperses the image of a 
point source into a short spectrum \citep{filippenko82}, so an 
important proviso to the use of this simple form of $\Omega$ is 
that the integrals are only applicable to a \emph{single 
wavelength\/} for all observations away from the zenith in the 
absence of atmospheric dispersion compensation.  Chromatic 
aberration in the telescope optics has similar 
wavelength-dependent effects on $\Omega$.  The problem 
atmospheric dispersion raises for spectrophotometry is not 
treated further here \citep[for a comprehensive analysis, 
see][]{donnellyetal89} but the related effects of atmospheric 
dispersion on \emph{guiding\/} and differential atmospheric 
refraction across the telescope field are addressed in 
\S\ref{sec:atmos-disp} and \S\ref{sec:diff-refrac}.  Although 
\eqn\ref{eq:eta_point_source} applies to any aperture, this paper 
deals with fibers which are modelled as simple circular apertures 
of diameter $d$.

\section{Sources of position errors}

The total position error for a fiber, $\mathbf{r}$, may be 
defined as the vector from the center of the physical fiber 
entrance aperture to the center of its target objects' image on 
the telescope focal surface, so $\mathbf{r}=0$ indicates an image 
centered and focused on the fiber entrance aperture.  The 
magnitude of $\mathbf{r}$ can be resolved into its projection 
onto the focal surface, \rn, which moves the fiber on a plane
orthogonal to the optical axis, and its projection onto the 
optical axis, \rp, which moves the fiber in a direction parallel 
to the optical axis.  In addition, the alignment between the axis 
of the fiber at its entrance aperture and the incident telescope 
beam must be considered. 

Sources of orthogonal, parallel and alignment angle errors will 
now be described in the approximate order in which each is 
introduced into observations.  Orthogonal errors are measured 
either by linear offsets across the focal surface or by angular 
offsets across the sky.  Parallel errors are measured by linear 
offsets along the telescope axis.  Alignment errors are measured 
by the telecentric angle, \teleang\ (\ie\ the angle between the 
fiber axis and the central ray of the telescope beam to the 
fiber).  Parameters used are $f$, the telescope focal ratio; $p$, 
the mean telescope image scale in the sense of angular separation 
on the sky per unit linear separation on the focal surface; 
$\delta{r}$, the contribution to a linear offset from one source 
of error;  $\theta$, the total angular displacement on the sky; 
$\delta\theta$, the contribution to $\theta$ from one source of 
error; \WF, the FWHM of a point-source image on the telescope 
focal surface; and \WA, the FWHM of the image of a point source 
measured at the fiber entrance aperture.  Airmass is measured by 
$\sec{z}$, where $z$ is the angular separation of a point on the 
sky from the zenith.  Unless otherwise noted, each source of 
error is independent of the others.  Orthogonal errors may thus 
be combined as $\theta^2 = (\rn\, p)^2 = \sum (\delta\theta)^2 = 
\sum (\delta\rn\, p)^2$, and parallel errors combined as $\rp^2 = 
\sum (\delta\rp)^2$.  

\subsection{Errors orthogonal to the optical axis}

\subsubsection{Astrometry}

The first source of orthogonal error comes from the astrometry in 
the target input catalog.  It is easy to suppose that recent 
all-sky astrometric catalogs should make it is simple to achieve 
\rms\ astrometric errors $\delta\theta \ll 1$~arcsec over wide 
fields, but this is not necessarily the case.  First, astrometric 
catalogs such as the USNO-A2.0 \citep{monetetal98} that combine 
faint magnitudes and all-sky coverage have in general been 
derived from measurements of single-epoch or averaged-epoch 
photographic survey plates with ages of up to 50 years or more, 
introducing large and variable proper motion uncertainties to 
positions at the current epoch.  

Even with astrometric catalogs that include good proper-motion 
data, such as Tycho-2 \citep{hogetal2000}, many other factors 
must be taken into account to achieve accurate calibration of 
target positions.  The next requirement is that the centroids of 
the astrometric standard stars and the target objects are found 
in the source material to good precision and in the same 
reference frame.  This is not trivial when wide magnitude ranges 
are covered, particularly for target catalogs based on 
photographic plates or that rely on brighter astrometric 
catalogs, as the reference stars may be saturated or comatic in 
deep source material.  Tycho-2, for example is complete at the 95 
per cent level only to $V\simeq11.5$, still much brighter than 
the guide stars and targets typical of fiber spectroscopy.  The 
forthcoming USNO-B catalog \citep{monet2000} and the 
second-generation Guide Star Catalog \citep{morrisonm2001} which 
will combine all-sky coverage, high-precision positions to faint 
magnitudes and accurate proper motions promise to be of great use 
for fiber spectroscopy.

Target positions derived from wide-field photographic plates may 
also include magnitude-dependent systematic errors of 
$\delta\theta\simeq0.05$~arcsec~mag$^{-1}$ \citep{luetal98}.  
Atmospheric dispersion can introduce a color-dependent shift in 
apparent positions in any source material as it does in 
spectrographic observations (\S\ref{sec:atmos-disp}), so it is 
important to match the filters used for the source material to 
the wavelengths for which spectra are to be obtained or to apply 
an appropriate correction.  

Where the input catalog is derived from tiled observations, for 
example from adjacent photographic survey plates, great care 
\citep[see for example][]{tafflb90} must be taken at the 
boundaries between plates to avoid steep local gradients or step 
inhomogeneities in the spatial distribution of astrometric errors 
in a target catalog which may otherwise have a small mean error 
on the scale of each spectrographic field \citep{newman99}.  
Similar care is needed when the astrometric calibration catalog 
is broken into separate zones, as with the Guide Star Catalog 
\citep{russelletal90}, or at the boundaries of the surveys from 
which the astrometric catalog was constructed.  The USNO-A2.0 
catalog, for example, has been shown to include discontinuities 
of $\sim0.3$~arcsec at the boundary between the northern and 
southern Schmidt surveys from which it was built 
\citep{assafinetal2001}.  

Note that a first-order \emph{systematic\/} astrometric error  
that shifts all fibers by the same vector is easily removed when 
the guide stars are centered in their fibers during observations, 
provided the guide star positions are in the same astrometric 
system as the target object positions.  If the fiber mount plate 
can be rotated on the telescope, then a systematic rotation (such 
as that from precession) can also be easily removed.

\subsubsection{Aberration, parallax and proper motion}

The motion of an observer relative to an observed object produces 
a tilt of the apparent direction of the object toward the 
direction of motion of the observer due to the finite speed of 
light.  This tilt is known as stellar aberration, or often simply 
aberration.  The magnitude of the tilt depends on the relative 
velocity and the angle, $\Theta_a$, between the directions of 
observation and motion.  The full correction naturally involves 
the special theory of relativity, but for terrestrial 
observations, the change in apparent position of an object due to 
the Earth's motion about the Sun is well approximated by
\begin{eqnarray}
  \Delta\alpha & = & \frac{-k}{\cos\delta}(\sin\Lambda \sin\alpha + 
  \cos\Lambda \cos\epsilon \cos\alpha)  \nonumber \\
    \Delta\delta & = & -k\sin\Lambda \cos\alpha \sin\delta \\
    & & + k\cos\Lambda(\cos\epsilon \sin\alpha \sin\delta  - \sin\epsilon \cos\delta) \nonumber
\end{eqnarray}
where $k = 20.496$~arcsec is the constant of aberration for the 
Earth's orbital motion about the Sun, $(\alpha, \delta)$ are the 
true right ascension and declination of the object, $\Lambda$ is 
the true solar longitude, $\epsilon$ is the obliquity of the 
ecliptic, and $(\Delta\alpha, \Delta\delta)$ are the change in 
apparent position in the sense apparent coordinate minus true 
coordinate \citep[and references therein]{lang86}.  

Annual aberration due to the Earth's orbital motion about the Sun 
ranges from $\Delta\Theta_a=0$ for $\Theta_a=0$\degr\ to 
$\Delta\Theta_a=20.47$~arcsec for $\Theta_a=90$\degr, so the mean 
\emph{differential\/} annual aberration across a telescope field 
is $20.47/90\simeq0.23$~arcsec~deg$^{-1}$ in the sense that the 
field is compressed in the direction parallel to the Earth's 
motion.  The annual aberration at the planned time of observation 
should therefore be included when deriving fiber positions to 
remove this differential displacement across the telescope 
field.  The maximum diurnal aberration due to the rotation of the 
Earth is $\sim 0.32$~arcsec, so the mean differential diurnal 
aberration is $\sim0.004$~arcsec~deg$^{-1}$ and may therefore be 
neglected.   

Because aberration depends on both $\Lambda$ and $\Theta_a$, the 
correction for annual aberration for a target field depends on 
the date of observation as well as the hour angle.  For example, 
observing a 3\degr-wide equatorial field one lunation later than 
that for which the fiber positions were designed, under the 
circumstances in table \ref{tab:circumstances}, results in a 
change in field compression due to differential aberration of 
$0.24$~arcsec.  Assuming the field is centered on the mount plate 
in both cases, images at the edge of the field are therefore 
displaced by $\delta\theta=0.12$~arcsec with respect to their 
nominal positions, with $\delta\theta$ for other fibers being 
proportional to their radial distance from the center of the 
field.  This may be partially compensated for by adjusting the 
telescope image scale.  

Parallax of guide stars (and nearby targets) should also be 
corrected for the time of observation, but this can be difficult 
when anonymous guide stars are used, as their parallax will be 
unknown.  Similarly, proper motion of guide stars or nearby 
targets between the time of observation of the source material 
and the time of the spectrographic observations should be 
corrected, especially if the age of the source material is high, 
but again, this is only possible if the proper motions are 
known.  However, using sufficient guide stars for each observed 
field should make the mean parallax and proper motion errors 
negligible.  Choosing guide stars with hot spectral types will 
further reduce both the mean parallax and proper motion, because 
they will be at a greater distance than cooler stars for a given 
apparent magnitude.  If the spectral types of guide stars are not 
known with precision, then using a single color, \eg\ $(B-V)$ or 
$(U-B)$, to select guide stars with blue colors may suffice, but 
as discussed in \S\ref{sec:atmos-disp}, it may be more important 
to match guide star colors to target object colors.

\subsubsection{Conversion to focal-surface coordinates}
\label{sec:focal-map}

Accurate conversion of apparent positions on the sky into focal 
surface coordinates depends on accurate mapping of the focal 
surface.  \citet{cudworthr91} showed that it can be difficult to 
determine $p$ for wide-field telescopes to better than 1 part in 
$10^4$ and to determine the distortion coefficients that describe 
how $p$ varies across the field of view to better than 1 part in 
$10^3$.  An error in $p$ of this scale at a field radius of 
$1\degr$ leads to position uncertainties of 
$\delta\theta=0.36$~arcsec.  Even where the distortion is well 
mapped, errors in defining the center of the distortion field 
alone will lead to systematic errors in fiber coordinates.  In 
the case of the CTIO 4-m telescope \citet{cudworthr91} showed 
that a 1~mm error in defining the center of the distortion field 
leads to position errors of $\delta\theta \simeq 1.4$~arcsec at a 
25~arcmin field radius.  Because distortion from centering errors 
will in general be anisotropic, its effect cannot be fully 
removed by simply centering the guide stars and adjusting the 
mean image scale.  

Any lateral chromatic aberration in the telescope optical design 
will vary the distortion pattern by moving the center of \OmegaF\ 
orthogonal to the optical axis for each wavelength, spreading 
each image radially with respect to the center of the focal 
surface.  This wavelength dependence should be included when 
designing plate coordinates, unless the optical design provides 
good correction.  In the case of the 2dF, the 4-element corrector 
design leaves a residual variation of $<1$~arcsec over the full 
range of spectrographic wavelengths \citep{lewisetal2002}, 
significantly less than \WF.  

In drilled-plate systems, the bending of the fiber mount plate 
required to match the telescope focal surface 
(\S\ref{sec:plate-shape}) will also introduce a change in the 
position of the fiber mounting holes orthogonal to the optical 
axis.  In the SDSS case, for example, this amounts to $\sim 
75$~\micron\ at the edge of the plates, which must also be 
corrected in the plate design.  

Note that there are also separate problems introduced by  
temporal variations in mean $p$ (\S\ref{sec:scale}) and imperfect 
optical collimation (\S\ref{sec:collimation}).

\subsubsection{Fiber mounting}

The next source of orthogonal error comes from the mechanical 
positioning of the fibers on a mounting plate that is fixed in or 
subsequently moved to the telescope's focus.   Most fiber 
spectrographs use blind positioning of the fibers on a mount 
plate, where each fiber is positioned without reference to the 
actual positions of the target object images during observation.  
Coherent fiber bundles to image guide stars are positioned in the 
same way.  The fibers are attached to the mount plate either 
directly, by fixing a fiber-carrying ``puck'' to the plate with a 
magnet or glue, or by drilling holes in the plate into which 
ferrules, each carrying a centered fiber or fiber bundle, are 
plugged.  A robot positions the fiber puck or a drill bit.  Where 
a single $(x, y)$ robot arm is used to position many pucks or 
holes, the robot typically has $\delta\rn\simeq10$--$15$~\micron\ 
precision.  

There have been several significant exceptions to the blind 
positioning approach.  The Argus type of design 
\citep{ingerson93}\ uses individual robot arms for each fiber 
arranged as ``anglers around a pond'' at the periphery of the 
mount plate.  This approach increases the speed with which a set 
of fibers may be positioned at the cost of increasing mechanical 
complexity.  Argus itself cleverly allows for active fiber 
positioning in which fast sampling of a large synthetic aperture 
over each target image is used to find the actual centroid of the 
target and then make fine adjustments to the fiber 
positions\footnote{Of course, this only works when the centroid 
of the image is the desired target.}, achieving residual radial 
position errors of $\delta\theta \simeq 0.2$~arcsec, but offering 
the possibility of frequent correction to all other sources of 
error \citep{lutziss90}.  The FLAIR \citep{watsonp94} and MOFOCS 
\citep{pettersson88} designs position fibers using a human robot: 
the observing astronomer glues the pucks directly onto the images 
of targets on a copy of a photographic plate taken with the same 
Schmidt telescope used for spectroscopy, using a travelling 
microscope to position each fiber over the image of its target 
object.  This method avoids the need to transform coordinates 
from the source material, and may be cheap to implement, but will 
introduce additional random and systematic errors that vary from 
plate to plate.  \citet{pettersson88} estimates this introduces 
errors $\delta\rn\lesssim10$~\micron, but the actual value will 
be difficult to measure and will no doubt vary between 
astronomers and with individual fatigue \citep{parkerw98}.  In 
addition, unless the spectra are observed under the same 
circumstances as the original plate was exposed, no corrections 
will be applied for aberration, parallax, proper motion, 
atmospheric dispersion or refraction.  

A further source of orthogonal fiber mounting error comes from 
the centering of the fibers in their carrier pucks or ferrules.  
Minimizing these errors demands strict manufacturing controls, 
especially for cartridge-based instruments such as the SDSS 
spectrograph, for which $\sim6000$ fibers have been fabricated.  
For all plug-plate instruments the subsequent centering of the 
ferrules in the plug-plate holes is also critical, as there must 
be a balance between the requirement that differences between the 
diameters of the ferrules and the plug-plate holes be small 
enough to ensure concentricity, but large enough to allow actual 
insertion and removal of the ferrule \citep{siegmundetal98}.  
Typical \rms\ radial errors for the SDSS fibers, for example, are 
$\delta\rn\simeq9$~\micron\ for the fiber-ferrule concentricity 
and $\delta\rn\simeq8$~\micron\ for the ferrule-hole 
concentricity (Russell Owen, private communication, 2001).

\subsubsection{Temporal variation in image scale}
\label{sec:scale}

The fiber mount plate in most cases is metal and thus has a 
significant thermal coefficient of expansion, so corrections 
should be applied when the mount-plate coordinates are determined 
to compensate for the expected difference in temperature between 
the time the plate is drilled or the fiber pucks are placed and 
the time of observation.  Any residual difference between the 
design and actual temperatures will introduce orthogonal position 
errors.  Changes in the temperature of the telescope may also 
affect the image scale by moving the focal surface with respect 
to the primary mirror due to expansion or contraction of the 
telescope structure.  These errors will be systematic and 
proportional to the radial distance of each fiber from the center 
of guiding.  

This problem is more acute for instruments using pre-drilled 
plates because the time between positioning and observation will 
usually be much longer than for puck-positioning instruments, 
making it harder to predict the temperature at observation.  The 
aluminum alloys typically used for pre-drilled plates have a 
higher coefficient of expansion ($\sim25\times10^{-6}$~K$^{-1}$) 
than the steel used for magnetic-puck instruments 
($\sim12\times10^{-6}$~K$^{-1}$) or the glass used for glued-puck 
instruments ($\la10\times10^{-6}$~K$^{-1}$).  For surveys using 
many plug plates, this problem is further exacerbated by the 
difficulties of scheduling when each plate is actually observed.  
Experience on the SDSS shows that the extreme 
design-to-observation difference in temperature may be as large 
as $\sim20$~K for a plate repeatedly observed over a wide range 
of dates, resulting in scale differences of $\Delta{p}/p \simeq 
5\times10^{-4}$.  At the outside of the 328~mm-radius SDSS plug 
plates, this gives $\delta\rn\simeq165$~\micron\ equivalent to 
$\delta\theta\simeq2.7$~arcsec.  In contrast, robotic 
puck-positioning instruments such as the 2dF have the advantage 
of positioning the fibers as late as $\sim1$~hr before 
observation onto a plate that is already close to the observation 
temperature, so except at the start of each night when the rate 
of change of temperature may be large, the temperature difference 
between the time of plugging and the time of observation is 
unlikely to $\ga2$--$3$~K.  In the case of the 2dF, this 
difference has been found to be small and quite predictable 
although as yet uncorrected \citep{lewisetal2002}.  

There are generally two options available to compensate for 
residual image scale errors at the time of observation for puck 
and plug-plate instruments.  The telescope may be defocused 
slightly, as this will often also change the effective image 
scale, but at the cost of a compromise between image scale and 
point-spread function size.  This may be the only adjustment 
practical for instruments mounted at prime focus.  Alternatively, 
in Cassegrain and other multiple-mirror systems, the position of 
the mount plate with respect to the primary mirror may be moved 
to change the effective focal length of the telescope and the 
secondary then moved to refocus.  The SDSS telescope achieves 
this by moving the primary mirror along the optical axis, and is 
able change its image scale by as much as $\Delta{p}/p \simeq 
\pm5\times10^{-4}$ while still being able to achieve optimal 
focus.  This permits the image scale error during exposures to be 
maintained to within a factor of $\Delta{p}/p \simeq 
\pm2\times10^{-5}$ of optimal (equivalent to 
$\delta\theta=0.1$~arcsec at the outside of its fiber mount 
plates) by applying scale corrections derived from automatic 
analysis of the observed positions of guide stars in their 
coherent fiber bundles.  The mount for the Fruit \& Fiber 
instrument on the du Pont 2.5m telescope at \LCO\ can similarly 
move with respect to the primary by $\sim3$~cm\ 
\citep{shectman93}, allowing for a change of $\Delta{p}/p \simeq 
\pm8\times10^{-5}$, although without the aid of automated 
analysis of scale error during observations.   

\subsubsection{Collimation and field rotation}
\label{sec:collimation}

Imperfect optical collimation introduces an optical distortion 
that remains static with respect to the optics, but which rotates 
during exposures with respect to the images, systematically 
moving the images with respect to the fibers.  Good collimation 
is therefore particularly important in telescopes with alt-az 
mounts where there may be a large rotation of the field during 
each exposure.  In the case of the SDSS telescope, this is 
expected to produce orthogonal errors $\delta\rn \la 10$~\micron\ 
\citep{knappetal99} equivalent to $\delta\theta \la 
0.17$~arcsec.  Because field rotation is much smaller on 
equatorially-mounted telescopes, instruments such as 2dF incur 
essentially zero variable error due to rotating distortion, but 
may still suffer from fixed-pattern errors if collimation is not 
perfect (\S\ref{sec:focal-map}).  

\subsubsection{Atmospheric dispersion and guiding}
\label{sec:atmos-disp}

For a plane-parallel approximation of the atmosphere applicable 
at zenith angles typical of astronomy other than for solar 
observations near the horizon, the angle of refraction due to the
atmosphere is given by 
\begin{equation}\label{eq:atmos_refrac}
  \Rlz = (n_\lambda - 1) \tan{z}
\end{equation}
where $n_\lambda$ is the refractive index of the atmosphere at 
wavelength $\lambda$ relative to the vacuum and $z$ is again the 
zenith angle \citep{green85}.  \citet[and references therein]
{filippenko82} shows how $n_\lambda$ varies with atmospheric 
conditions and how \Rlz\ consequently varies with observing 
circumstances.  

The first effect of \Rlz\ on fiber positions comes from its 
dependence on $\lambda$, so that images of point sources observed 
away from the zenith are dispersed into short spectra orthogonal 
to the horizon, with the angular distance between two wavelengths 
$\delta\theta=R(\lambda_1,z) - R(\lambda_2,z)$.  For example, at 
the center of the field under the designed observing 
circumstances in table \ref{tab:circumstances}, 
$\delta\theta\simeq1.23$~arcsec.  The effects of this atmospheric 
dispersion on spectrophotometric measurements with fiber 
spectrographs and the measures necessary to correct them are not 
treated here; for a comprehensive analysis, see 
\citet{donnellyetal89} and \citet{filippenko82}.  However, fiber 
positions should be determined for a specific $\lambda$.  This 
may be a common wavelength for all fibers, typically close to the 
central wavelength recorded by the spectrographs, or it may vary 
between target object types to record the most scientifically 
interesting regions of the target spectra.  

Atmospheric dispersion must also be accounted for when there is a 
difference between the wavelength for which the fiber positions 
are designed and the apparent SED of the guide stars as recorded 
by the guide camera.  This is particularly important when 
observing intrinsically-blue targets such as quasars, as most 
guide stars will have an apparent SED with a peak at a much 
longer wavelength, which will therefore appear at a different 
initial offset within the dispersed images and see a different 
variation in offset with $z$.  The problem is further increased 
because guiding normally tracks the brightest part of each guide 
star image, and the peak of the guide star SED varies from star 
to star and in any case is generally unknown.  This problem may 
be reduced by filtering the guide camera.  Refraction increases 
rapidly as $\lambda$ decreases, so it is better to guide on the 
position of the red end of the dispersed guide-star images 
because a broad filter may then be used to admit sufficient light 
to the guide camera without incurring large differences between 
the position of the peak of the filtered guide star SED and the 
filter central wavelength for a variety of guide star SEDs.   For 
example, guiding at $\sec{z} \le 1.6$ through a filter with a 
central wavelength $\lambda_c=6000$~\AA\ and a bandpass 
$\Delta\lambda=2000$~\AA\ results in position errors of 
$\delta\theta\le0.05$~arcsec for guide stars with temperatures 
between 3000~K and 30000~K \citep{cuby94}.  An offset must then 
be applied between the guide and target fiber positions to place 
the optimum \emph{observed\/} wavelength of the target spectra 
onto their fibers while tracking the filtered guide stars, but 
this offset increases the dependence of the fiber positions on 
the designed zenith angle.   

It is theoretically possible to design an atmospheric dispersion 
compensator (ADC) by introducing prismatic elements into the 
optical path to cancel the effects of atmospheric dispersion 
without introducing other aberrations.  The design and 
construction of precision ADC optics, as has been done for the 
2dF spectrograph, is both complicated and costly 
\citep{lewisetal2002}.  In order to work at different airmasses, 
the prismatic elements must change their relative positions.  
This requires that they be collimated to very high precision or 
they will introduce further complications of image motion during 
long exposures \citep{willstrop87}.  Because of these 
complications, many fiber spectrographs dispense with ADCs and 
instead rely on larger fibers to collect the dispersed image.

\subsubsection{Atmospheric differential refraction}
\label{sec:diff-refrac}


The second effect of \Rlz\ in \eqn\ \ref{eq:atmos_refrac} comes 
from its dependence on $z$.  For any non-zero field width, 
observations centered at $z=0$ will be isotropically compressed 
by the difference in refraction between the zenith and the edge 
of the field.  As the observation moves away from the zenith, the 
compression parallel to the horizon remains practically constant 
but that orthogonal to the horizon increases.  For example, a 
$3\degr$ field at the designed circumstances in table 
\ref{tab:circumstances} will be observed at $z=35.7\degr$ and 
compressed vertically by $3.5$~arcsec at the central wavelength.  
When determining fiber mount-plate coordinates, a correction for 
\Rlz\ must therefore be included, but problems arise if either 
the exposure is long enough for a significant change in airmass, 
or the actual observation is centered at an airmass significantly 
different from that for which the fiber positions were 
determined.  If the same field is observed just $40$~min late, 
then $z=40.4\degr$ and the field compression increases to 
$4.0$~arcsec.  Assuming the field is centered on the mount plate 
in both cases, images at the top and bottom edges of the field 
will be displaced by $\delta\theta=0.25$~arcsec by an hour angle 
change of only $10\degr$.

For fields at high altitude or designed for observation away from 
the meridian, a partial compensation for the change in field 
compression between the designed and observed $z$ is possible by 
changing the image scale of the telescope by
\begin{equation}\label{eq:compress}
  \frac{\Delta{p}}{p} = 
  \frac{(n_\lambda - 1)\,(\tan^2{z_{\rm{des}}} - \tan^2{z_{\rm{obs}}})}{2}
\end{equation}
where $z_{\rm{des}}$ and $z_{\rm{obs}}$ are the designed and 
observed zenith angles, and the division by $2$ equalizes  the 
residual $\delta\theta$ error for a circular telescope field so 
that the movement of fibers towards the images perpendicular to 
the horizon is balanced by their movement away from the images 
parallel to the horizon.  However, for fields at low altitude 
observed near the meridian, the limiting factor in determining 
the acceptable range of hour angles is the conflict between field 
rotation and fibers positioned for a central field compression, 
for which a combination of pointing, plate rotation and a scale 
change according to \eqn\ \ref{eq:compress} provides only partial 
correction.  The uncorrected pattern will be a quadrupole with an 
amplitude proportional to the radial distance from the center of 
the plate.

\subsubsection{Pointing and instrument rotation}
\label{sec:guiding}

The final source of orthogonal errors appears during the 
exposures.  Random variations in telescope pointing about a 
constant mean position on the sky with variation time scales much 
shorter than the exposure time, when integrated over the 
exposure, will have the effect of convolving \OmegaF\ with a 
Gaussian of width defined by the variance of the pointing 
errors.  However, such pointing errors are \emph{not\/} 
necessarily random in distribution and their correction will 
typically occur only a few times per minute depending on the 
integration time of the guide camera.  Pointing errors may thus 
introduce a mean orthogonal fiber positioning error affecting all 
fibers equally.  A reasonable expectation for modern automated 
guiding systems using multiple guide stars is that the \rms\ 
error will be $\delta\theta\simeq0.1$~arcsec.  

A mean error in the fiber mount plate rotation angle, 
$\delta\Phi$, will introduce an orthogonal error $\delta\theta = 
p\,r\,\delta\Phi$ where $r$ is the distance of the fiber from the 
center of rotation.  However, this effect is generally very 
small.  In the case of SDSS observations, for example, it is 
normal to maintain $\delta\Phi \la 2$~arcsec, producing 
$\delta\theta\la0.05$~arcsec at the edge of the 328mm radius 
mount plates with an image scale $p=16.5$~arcsec~mm$^{-1}$.

\subsection{Error parallel to the optical axis}

\subsubsection{Shape of the fiber mount plate}
\label{sec:plate-shape}

Without corrector optics, the focal surface of a telescope is 
unlikely to be close enough to a plane to permit the fiber mount 
plate to be planar without introducing significant variations in 
image quality across the field of view.  In addition, the focal 
surface may vary with wavelength due to longitudinal chromatic 
aberration in the telescope optics bringing different wavelengths 
to a focus at different positions parallel to the optical axis.  

In principle, the fiber mount plate may be deformed when mounted 
on the telescope in order to bring all the fibers to the focus of 
the telescope beam, at least for some specific wavelength.  The 
amount of distortion of the mount plate required to follow the 
focal surface depends on the optical design of the telescope and 
correctors and can vary quite widely between telescopes.  For 
example, the sagittal depth (\ie\ the depression of the center of 
the focal surface below the plane defined by the periphery of the 
field) of the SDSS telescope with the spectrographic corrector in 
place is 2.6~mm \citep{knappetal99}, while that of the 2.5-m du 
Pont telescope is 7~mm \citep{shectman93}.  For drilled-plate 
systems, bending of the fiber mount plate is usually applied when 
the plate is attached to the telescope or installed in its 
cartridge, as in the Fruit \& Fiber instrument, which distorts 
the plate by loading the edges of the plate with a pneumatic 
press, or the SDSS fiber cartridges, which use a combination of 
bolt-down edge loading and a central pull-down bar to achieve an 
acceptable profile.  In practice, however, the deformed fiber 
mount plate is unlikely to follow the true shape of the focal 
surface, leading to defocusing errors which can vary also widely 
between instruments.  The typical deviation of an SDSS fiber 
mount plate from the true focal surface measured at \APO\ when 
the plate is in the optimum position is 
$\delta\rp\simeq80~\micron$, which in an $f/5$ beam convolves 
\OmegaF\ with a top-hat profile of diameter equivalent to $\sim 
0.26$~arcsec on the sky.  
For the Fruit \& Fiber spectrograph, the maximum deviation is 
typically $\delta\rp\simeq1000$~\micron\ \citep{shectman93}, 
equivalent to convolution of \OmegaF\ with a top hat of $\sim 
1.4$~arcsec diameter, leading to a very large variation in image 
sizes across the field.  

For magnetic-puck instruments, the required distortion may be 
achieved by permanent bending of the mount plate, although this 
may complicate the requirements for the robotic positioner.  
Instead, the 2dF uses a corrector design that achieves a 
remarkably flat focal surface that is within $\delta\rp \le 
30$~\micron\ of planar across its entire field 
\citep{lewisetal2002}.  The maximum displacement is equivalent to 
the convolution of \OmegaF\ with a top-hat profile of just 
$0.13$~arcsec diameter, which is small compared to the mean 
seeing.  

Further small parallel errors will result from fiber to fiber 
variations of the position of the fiber aperture with respect to 
the mounting surface of the ferrule or puck.  Manufacturing 
tolerances can be expected to be of the same order as those for 
other ferrule dimensions, giving $\delta\rp\simeq5$~\micron.  For 
plug-plate instruments there will be an additional and 
potentially larger variation in the actual seated position of the 
ferrule in the plug plate.  Where the fibers are plugged into a 
mount plate already attached to the telescope, as with the Fruit 
\& Fiber instrument, careful manipulation should ensure the 
seating error is negligible.  However, experience has shown 
seating errors to be a problem for the SDSS plug-plate cartridges 
which are transported on a cart between the support building, 
where fibers are plugged into the mount plate, and the 
telescope.  Vibration during this transport has on rare occasion 
caused a few fibers to fall completely out of their holes, and is 
thus likely to also move other fibers away from their 
fully-seated positions.  At the time of writing, the resulting 
distribution of \rp\ is unknown, but it is suspected that holes 
near the periphery of the plates will be less prone to errors 
than those near the center as the peripheral fiber carriers for 
those holes tend to be more tightly bent, thus putting larger 
torque on the ferrule-hole surface and so more firmly securing 
the ferrule.  For instruments that glue pucks to the mount plate, 
variations in the thickness of the glue used to attach the fiber 
to the plate will vary \rp, but with careful application this 
also should be a matter of only a few microns.  
Magnetically-attached pucks should only incur the constant 
manufacturing error per puck, provided all mating surfaces are 
kept clean and free of corrosion.

\subsubsection{Focus errors during observation}

In Cassegrain systems, the secondary mirror position required to 
maintain focus will vary during observations due to changes in 
the temperature of the telescope and flexure in its structure.  
Similarly, the focal adjustment will vary with telescope 
temperature in prime-focus systems.  Detecting focus variations 
requires a measurable image of the focal surface, typically from 
the coherent fiber bundles used to image the guide stars.  Focus 
corrections are therefore subject to the same sampling and 
discreteness problems as guiding corrections (see 
\S\ref{sec:guiding}), but also suffer from the additional problem 
that dynamic variations in the size of the true seeing disk, \WF, 
due to atmospheric changes may mask focus errors.  Nevertheless, 
skilled observers should be able to maintain the images to within 
$(\WA - \WF) \simeq 0.2$~arcsec of best focus, equivalent to 
$\delta\rp = (\WA - \WF)\,f/p$~\micron.  Systems to automate 
continuous focus adjustment may be possible if guide fibers can 
be displaced above and below the optimum focal surface to provide 
measures of differential focus.  Note that corrections to focus 
are commonly accompanied by changes in image scale (see 
\S\ref{sec:scale}).

\subsection{Sources of alignment error} \label{sec:angular}

Light may be lost between the exit aperture of the fibers and the 
dispersing element of the spectrograph if the axis of the fiber 
at its entrance aperture is not aligned with the central ray of 
the incident light beam in the so-called ``telecentric'' position 
\citep[and references therein]{wynne92}.  This alignment may 
differ considerably from normals to the focal surface.  

Drilled plug-plate systems offer a distinct advantage over 
puck-mounted fibers in this respect because the holes in the plug 
plate can be drilled to tilt all of the fibers close to 
$\teleang=0$ for any fiber arrangement.  For example, in the SDSS 
spectrograph the telecentric angle of normals to the focal 
surface may be as large as $\teleang \simeq 1.84\degr$.  The SDSS 
plates are deformed while being drilled such that when 
subsequently deformed (to a different shape) to follow the focal 
surface, the \rms\ fiber alignment error is $\teleang < 0.2\degr$ 
\citep{siegmundetal98}.  However, the bending during drilling 
introduces yet another radial displacement of the holes that must 
be corrected in the plate design.  

Puck mounts, on the other hand, must be usable over a wide range 
of positions and orientations, and so must be designed to accept 
light at an angle to the usually flat mounting surface that is a 
compromise to the variation in \teleang\ over the area each puck 
reaches, potentially resulting in larger losses.  The 2dF mount 
plate, for example, has $0\degr \la \teleang \la 4\degr$.  Its 
pucks have a $2\degr$ offset built into their input prisms, 
leaving an uncorrected fiber alignment error of up to 
$\teleang\simeq2\degr$ \citep{lewisetal2002}, or $\sim13$ per 
cent of the $f/3.5$ corrected input beam width.

\section{Efficiencies and aperture corrections}
\label{sec:efficiency}

Increasing \rn\ will decrease $\eta$ for point sources because 
the fiber aperture will clearly be illuminated by a fainter 
region of the image\footnote{For extended sources, increasing 
\rn\ may decrease or increase the total signal recorded, 
depending on the nature of the object being observed, but if it 
increases the signal, it will be the \emph{wrong\/} signal coming 
from an part of the object other than the intended target.  }.  
Increasing \rp\ for point sources will, in general, also decrease 
$\eta$ because the defocused image illuminating the fiber 
aperture will have a lower surface brightness.  The detailed 
transformation of \OmegaF\ into \OmegaA\ by defocusing depends on 
the nature of the telescope optical configuration, in particular 
the central obstruction in reflecting telescopes 
\citep{schroeder87} which eventually produces a classical 
``donut'' image.  The defocused image typically has complex 
structure from imperfections in the figures of the mirrors and 
their collimation, which may include bright spots.  However, the 
effect of the small values of \rp\ that are likely during science 
exposures may be approximated to first order for point sources as 
a convolution of \OmegaF\ with a top-hat of angular diameter 
$p\,\rp / f$.  For practical purposes this may often be treated a 
simple increase in \WF\ by an additive term, giving the observed 
image FWHM at the aperture of $\WA \simeq \WF + (p\,\rp / f)$.  

\figs\ \ref{fig:eta}\ and \ref{fig:apcorr} show the fractional 
efficiency, $\eta$, and the equivalent loss in magnitudes, $m_a$, 
as a function of the total orthogonal error for a range of values 
of the total image FWHM measured at the fiber aperture.  The 
orthogonal error and image FWHM are both expressed in fiber 
diameters, \ie\ $\rn/d$ and $\WA/d$, so the plots are applicable 
to all fiber spectrographs or indeed other spectrographs with 
circular apertures.   

For a given survey, the actual orthogonal and parallel errors 
expected from each of the sources described above obviously 
depend upon many factors related to the sources of the target 
input catalog, the instrument and telescope design and the 
observational circumstances, so no attempt is made here to derive 
``typical'' total values for \rn\ and \WA.  However, once total 
values have been calculated for particular observational 
circumstances, \figs\ \ref{fig:eta} and \ref{fig:apcorr} can be 
used to determine the fraction of the light that will enter the 
fiber targeted on each object and the corresponding aperture 
correction.  

If fibers are not aligned in the telecentric position then $\eta$ 
is reduced by a further factor.  If $u$ is the angular radius of 
the telescope beam, $\theta_T$ is the telecentric angle and $B$ 
the fraction of the telescope beam that is unobstructed, then 
\citet{wynne92} has shown that the proportional light loss factor 
is
\begin{equation}
\gamma = 1 - \frac{k\theta_T}{Bu}
\end{equation}
where $k=0.6$ is accurate to within a few per cent for 
$\theta_T/u \la 1.4$.  The net efficiency is then $\eta = \gamma 
\eta_0$ where $\eta_0$ is the efficiency arising from fiber 
position errors in the telecentric condition.  Unless the 
spectrograph camera optics incorporate careful baffling after the 
fiber output, the lost light will become stray light in the 
spectrographs, further degrading the SNR.

\section{Conclusions}
\label{sec:conclusions}

This paper has quantified the principal factors that reduce the 
efficiency of observations made with fiber-input multi-object 
spectrographs.  These include:

\begin{itemize}
  \item Astrometry of the input catalog.  
  \item Corrections for parallax and proper motion of guide stars.  
  \item Corrections for annual aberration and atmospheric 
  refraction.  
  \item Optical collimation of the telescope and mapping of its image scale and distortion.  
  \item Precision of the fiber positioning system.   
  \item Curvature of the focal surface and the telecentric angle of the fibers.  
  \item Differences between designed and actual observing circumstances.  
  \item Precision of telescope pointing, instrument rotation 
  and focus during integration.  
\end{itemize}

\begin{figure}
 \psfig{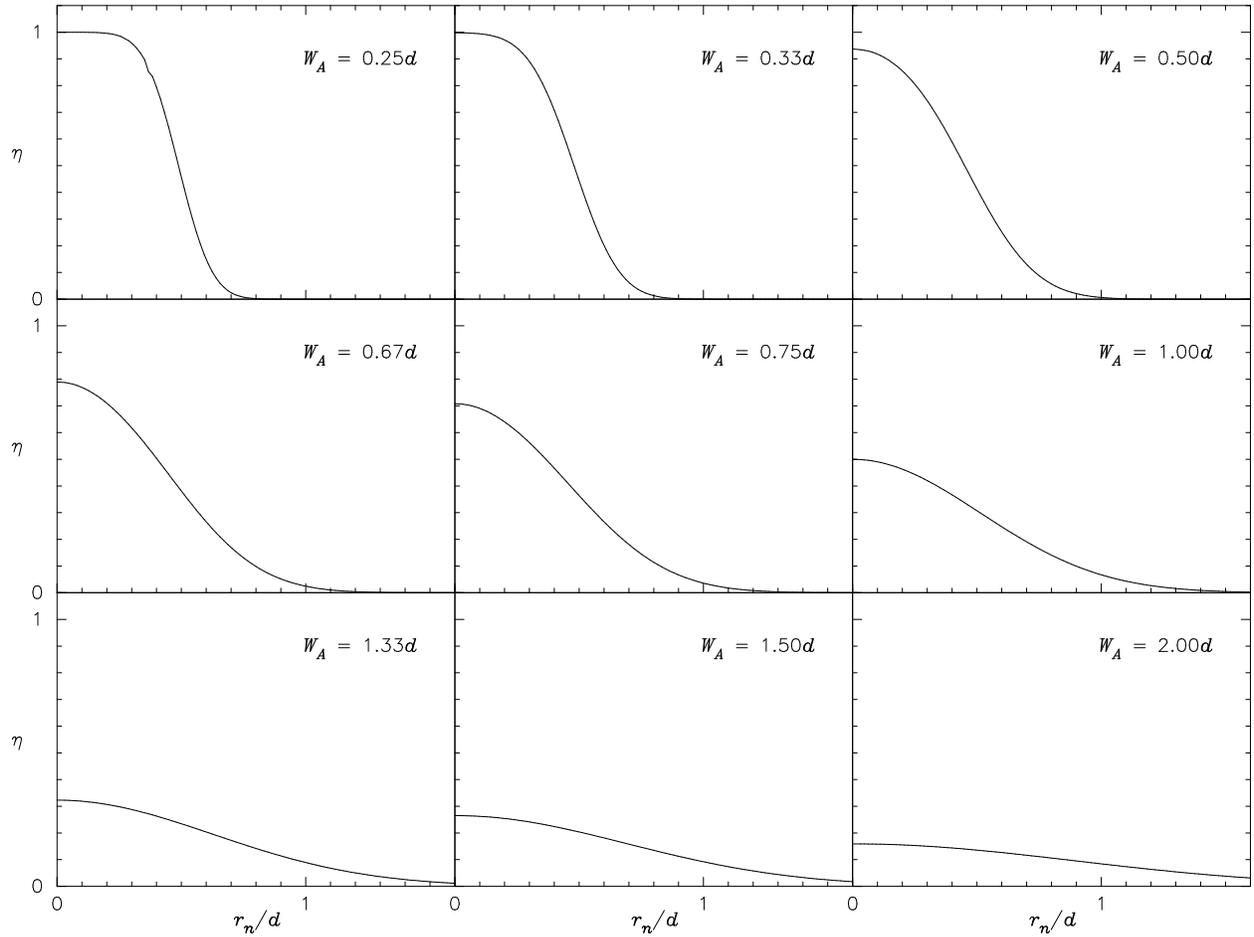}
 \figcaption{
 Fractional efficiency for point sources as a function of fiber-image displacement for a range 
 of image sizes.
 \label{fig:eta}
 }
\end{figure}

\begin{figure}
 \psfig{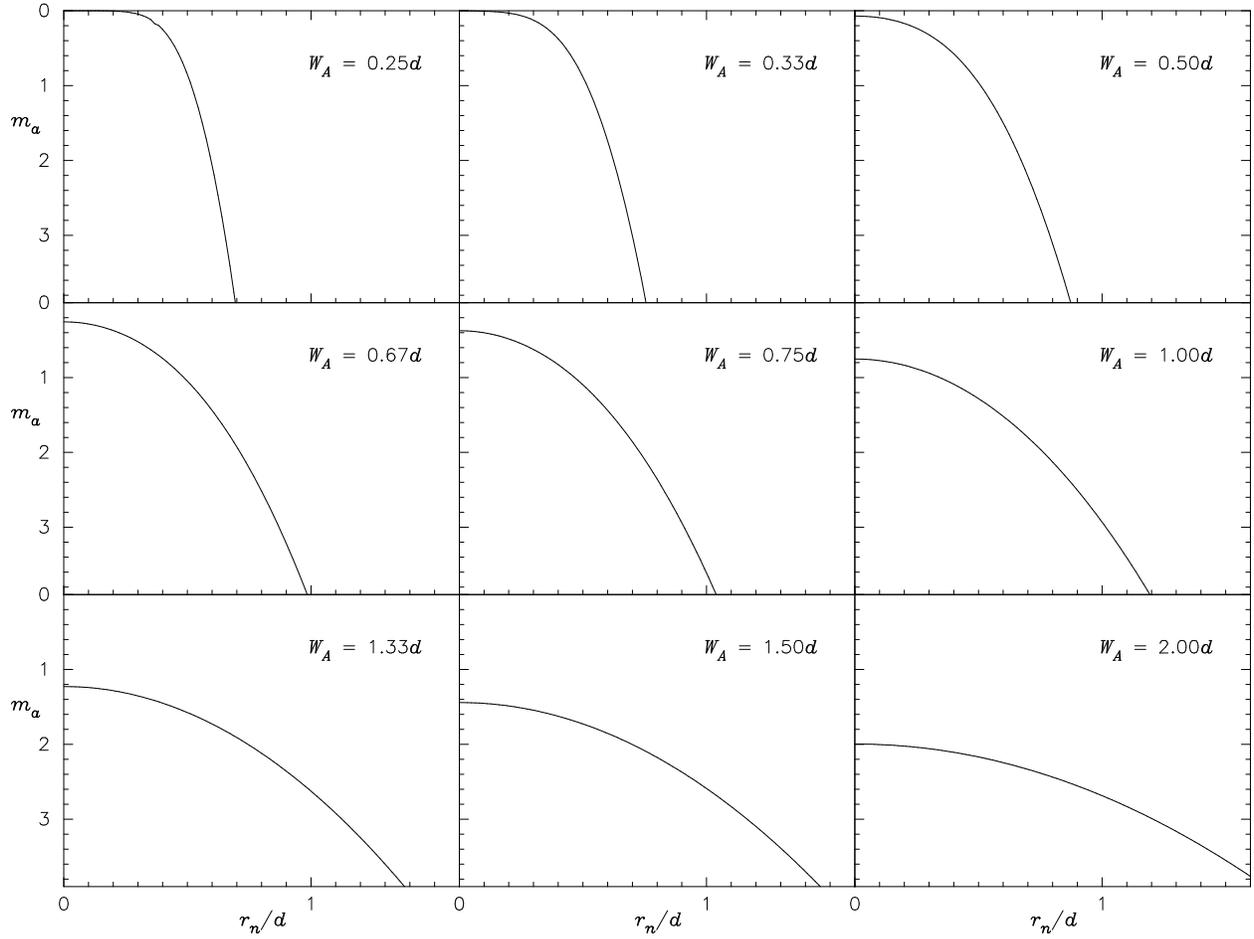}
 \figcaption[Aperture corrections]{
 Efficiency for point sources expressed as an aperture correction in 
magnitudes for the same fiber-image displacements and image sizes 
shown in \fig\ \ref{fig:eta}.
 \label{fig:apcorr}
 }
\end{figure}

Although small fibers may be desirable to reduce sky noise, 
\figs\ \ref{fig:eta}\ and \ref{fig:apcorr}\ show that small fiber 
diameters place tremendous demands on fiber positioning 
precision.  In the absence of ADC optics, small fibers also incur 
a heavier spectrophotometric penalty from atmospheric dispersion 
than larger fibers.  Poor seeing, which may be defined as $\WF 
\ga d$, greatly reduces the signal even from well-centered 
images.  Systems to automate continuous focus adjustment may 
clearly improve efficiency by minimizing $(\WA - \WF)$.  

Where tiled observations are used to construct a large survey, 
knowledge of the \emph{distribution\/} of fiber position errors 
across each field is as important as knowing the mean error or 
else the survey may suffer from a very complicated 
position-dependent faint-object selection effects, or may show 
errors in spectral identification or redshift measurement 
correlated with position.  In particular, when the fiber position 
error is proportional to the radial distance of the fiber from 
the field center because the telescope image scale is incorrect, 
the selection bias or spectral measurement errors may echo the 
tessellation in the survey catalog.  Removal of these effects for 
the analysis of large-scale structure requires very careful 
modelling.  

The individual observations for large tiled surveys will 
typically be made under quite variable conditions.  Even under 
photometric conditions, compensating changes to the exposure 
times used will be necessary when actual circumstances differ 
from those for which fiber positions were designed.  Minimizing 
the effects of atmospheric refraction demands that fields be 
observed over a small range of airmass with exposures centered on 
the meridian, but this forces a compromise between correcting the 
compression of a field and its rotation during exposure.  In 
practice, this constraint may cause great difficulties in 
scheduling observations around weather, particularly for 
instruments with fibers mounted well ahead of observation.  

While extra exposures will not remove systematic differences in 
SNR across each tile, they can at least ensure that the required 
survey minima are achieved for an acceptable fraction of all 
objects.  Nevertheless, additional observations beyond the design 
time for the fiber positions can leave strong 
\emph{spectrophotometric\/} effects in the data as the 
atmospheric dispersion changes which must be corrected by other 
means, such as synthetic large-aperture spectra of bright stars 
of known spectral type obtained by moving the telescope about the 
mean pointing, from which an approximate spectrophotometric 
correction can be derived.  

Given that the practical aspects of fiber positioning are fixed 
by the time the observations are made, what tools should 
observers have to maximize the efficiency of fiber spectrographic 
survey operations?  The first tool required to compensate for 
variations in fiber-to-image position errors is accurate 
real-time evaluation of the actual SNR values recorded in an 
exposure.  This will facilitate decisions as to how much 
additional exposure time is needed to achieve, but not 
excessively exceed, the required minimum SNR before fibers are 
reconfigured.  In addition to all-fiber mean values, such tools 
should provide information on the spatial distribution of SNR 
across each field so that position-dependent systematics can be 
reduced.  The SDSS observers at \APO\ have an excellent example 
of this in the form of a cut-down version of the full 
spectrographic reduction pipeline \citep{stoughtonetal2002} known 
as Son of Spectro (SoS; David Schlegel \& Scott Burles, private 
communications, 2001).  SoS runs on a fast dual-processor 
computer on the mountain and provides SNR totals and 
distributions together with other quality assurance parameters 
within minutes of the end of each exposure, enabling timely 
decisions by the observers on what exposures are required to 
complete a field.  

The importance of accurate and precise guiding in minimizing \rn\ 
and \rp\ demands a good auto-guiding system controlling all axes 
and other controls such as image scale and telescope focus.  
Efficient scheduling of observations with fiber spectrographs 
also requires a tool to help decide which field to observe next, 
in order to make best use of the available sky and conditions 
(for example, in avoiding moonlight).  This is particularly true 
for surveys using plug-plates, where the queue of fields 
available each night must be specified well in advance, and each 
plate is only suitable for observations over a particular range 
of dates and airmasses.  Again, the SDSS has a tool built around 
the fiber mount plate database specifically for this purpose 
(Stephen Kent, private communications, 2002).  This gives the 
observers up-to-date information on the plates available, the 
results of prior observations for the currently plugged plates, 
the sky brightness due to moonlight in the direction of each 
spectrographic field, and the optimum and allowable times of 
observation for each plate.  Other factors not related to 
observational circumstances may also be included in such a tool, 
for example to record the scheduling priority assigned to each 
field when aiming to achieve timely contiguous coverage of large 
tiled areas.  

Finally, it should be stressed that while the full extent of the 
problems of positioning fibers for efficient observation may seem 
daunting to new fiber-spectrographic observers, surveys such as 
the SDSS and 2dF that are relying on fiber spectrographs are 
achieving tremendous successes, clearly demonstrating that the 
problems can be adequately solved by attention to detail and the 
application of good engineering practice.  The work required to 
achieve similar success in smaller scale observational programs 
with fiber spectrographs is just as difficult, so it is to be 
hoped that the tools developed for the large surveys will become 
available for broader use in due course.

\acknowledgments

I thank Terry Bridges, Russell Owen, and Jim Gunn for useful 
discussions, Steve Kent for a critical and very helpful review of 
a draft manuscript, and the anonymous referee for constructive 
comments.

\clearpage
 
\clearpage

\begin{deluxetable}{lcc}
 \tablecaption{Hypothetical observational circumstances used 
 for examples, based on an equatorial field observed with the 
 SDSS spectrographs with physical conditions typical of winter observing 
 at \APO.  The designed and actual dates of observation differ by one 
 lunation.  
 \label{tab:circumstances}
 }
\tablewidth{0pt} 
\tablehead{
 \colhead{Parameter} &
 \colhead{Actual} & 
 \colhead{Designed}
} 

\startdata

Telescope focal ratio $f$               &  5  & \\
Field width                             & $3\degr$  & \\
Image scale $p$                         & 16.5~arcsec~mm$^{-1}$ & \\
Recorded wavelength range               & $3800$--$9200$~\AA & \\
Observatory altitude above sea level    & 2788~m & \\
Air pressure                            & 550~mm Hg & \\
Ambient temperature                     & $-4$~C & \\
Relative humidity                       & $40$\% & \\
Field center right ascension            & $0\rah$ & \\
Field center declination                & $0\degr$ & \\
Central hour angle                      & $25\degr$     & $15\degr$ \\
Date of observation                     & 2002 Feb 1    & 2002 Jan 3 \\

\enddata
\end{deluxetable}

\end{document}